\documentstyle[twocolumn,aps,psfig]{revtex}
\begin{document} 
\twocolumn[\hsize\textwidth\columnwidth\hsize\csname @twocolumnfalse\endcsname
\draft
\title{Quantum antiferromagnetism in the $d=3$ Hubbard model --- \\
a spin-fluctuation approach} 
\author{Avinash Singh\cite{asingh}} 
\address{Theoretische Physik III, Elektronische Korrelationen und Magnetismus,
Universit\"{a}t Augsburg, D-86135 Augsburg, Germany}
\maketitle
\begin{abstract} 
A self-consistent spin-fluctuation theory is developed 
to obtain $T_{\rm N}$ vs. $U$ for the half-filled 
Hubbard model in the whole $U/t$ range. 
Good agreement is obtained in the strong coupling limit
with the high-temperature series-expansion result for the
equivalent Heisenberg model.
Quantum spin-fluctuation correction to the sublattice magnetization
is also obtained for all $U$ at the one-loop level.
A spin picture is used throughout, and quantum effects are 
incorporated through transverse spin fluctuations, which 
are evaluated in the random phase approximation
using a new method. 

\end{abstract} 
\vskip2pc]

Finite-temperature antiferromagnetism 
within the three-dimensional Hubbard model at half filling
involves two fundamentally different aspects. 
One is the local symmetry breaking at temperatures much smaller than 
the AF gap parameter $\Delta$, 
and formation of the local-moment 
which grows rapidly and saturates with increasing $U/t$. 
The other is the scale of low-energy spin excitations, 
which initially rises along with the charge gap for small $U$,
and then crosses over to an approximately $t^2/U$ fall-off for large $U$,
becoming degenerate, in the strong coupling limit,
with the exchange energy scale $J=4t^2/U$ 
of nearest-neighbor (NN) spin couplings in the equivalent 
$S=1/2$ quantum Heisenberg model.
Thus, while in the weak-coupling limit ($U<<B=12t$, the free-particle band-width)
it is essentially the process of {\em moment melting}
which determines the magnetic transition temperature $T_{\rm N}\sim \Delta$, 
in the strong coupling limit  ($U>>B$)
it is the {\em spin disordering} tendency of
thermal fluctuations which determines the transition temperature 
$T_{\rm N}\sim J$ at which long-range AF order is destroyed.
A proper description of the magnetic transition in the whole $U/t$ range
within a {\em single} theory is therefore quite challenging.

Earlier studies of finite-temperature antiferromagnetism within the three-dimensional
Hubbard model, aimed at determining $T_{\rm N}$ vs. $U$ in the whole $U/t$ range, 
have employed the functional integral formalism,
mainly within the static 
approximation,\cite{hasegawa,kakehashi,kakehashi2,hasegawa2,moriya,moriya2}
quantum Monte Carlo methods,\cite{hirsch,scalettar}
and recently the dynamical mean-field theory,\cite{jarrell}
which becomes exact in the limit of infinite dimensions.\cite{vollhardt} 
All share the common feature of yielding a $T_{\rm N}$ which approaches,
in the strong coupling limit, 
the mean-field-theory (MFT) result 
for the equivalent spin-1/2 Heisenberg model, 
$T_{\rm N}^{\rm MF}=zt^2/U$, where $z=2d$ is the lattice coordination number.\cite{comparison}
While this is natural for the functional-integral schemes employing static approximation
and the dynamical mean-field theory, due to neglect of the Goldstone mode,
high-temperature series expansions yield a substantially lower 
$T_{\rm N} \approx 3.83t^2/U$.\cite{kakehashi2,hasegawa2,hirsch,scalettar}
A recent mean-field approach utilizing the Onsager reaction-field correction 
yields a $T_{\rm N}$ in close agreement.\cite{logan}
Here account is made of the low-energy spin excitations 
through an approximate mapping to an effective $S=1/2$ Heisenberg model with 
$U$-dependent, extended-range spin couplings.

In this paper we have considered a spin picture throughout
for the description of quantum antiferromagnetism in
the Hubbard model. We imagine quantum spins on each site,
and incorporate (quantum) transverse spin fluctuations about the
(classical) HF state, having local moment $\langle S_z\rangle_{\rm HF}$. 
Within this conceptually simple framework
we have obtained both the N\'{e}el temperature $T_{\rm N}$ vs. $U$ 
and the sublattice magnetization $m$ vs. $U$
(at $T=0$) in the whole $U/t$ range.
Transverse spin fluctuations are obtained 
within the random phase approximation (RPA)
in terms of magnon mode energies and amplitudes.
A new method is described for evaluating
the transverse spin correlations for arbitrary $U/t$. 

As $\langle S_z\rangle_{\rm HF}$ is the maximum (classical)
spin polarization in the z-direction,
and therefore also the maximum eigenvalue of the local $S_z$ operator,
hence $S\equiv \langle S_z\rangle_{\rm HF}$ plays the role
of the effective spin quantum number. 
At $T=0$, quantum corrections to the sublattice magnetization 
are then obtained at the one-loop level 
from the RPA-level transverse spin fluctuations.
On the other hand, at finite temperature a self-consistent theory 
is used in which the magnon spectral functions and energies are
self-consistently renormalized. Consistent with our RPA-level
description in which details beyond this level are ignored,
only momentum-independent multiplicative renormalizations are considered. 
While assumption is made here regarding their form,
these renormalizations are basic requirements 
following from the commutation relations of quantum spin
operators, and are further discussed later.

As all calculations are done with RPA-level magnon energies
and spectral functions, this 
provides a RPA-level description of the magnetic phase diagram.
Features such as continuous spin-rotational symmetry 
and transverse spin fluctuations are included,
resulting in substantial improvement over the HF theory.
Proper limiting results are therefore obtained 
as $d\rightarrow 2$ ($T_{\rm N}\rightarrow 0$, for all $U$),
and $d\rightarrow \infty$ ($T_{\rm N}\rightarrow T_{\rm N}^{\rm MF}$,
for $U/t\rightarrow \infty$).

We first discuss the approach for determining the N\'{e}el temperature
within the spin picture used here
in terms of the local spin operators $\vec{S}_i$.
In the anisotropic broken-symmetry state for $T<T_{\rm N}$, we have
$\langle S_i ^z S_i ^z \rangle > \langle \vec{S}_{i}^2\rangle /3$. 
However, with increasing temperature the anisotropy progressively decreases,
and eventually vanishes at the N\'{e}el temperature. Therefore we determine 
$T_{\rm N}$ from the following isotropy requirement,
\begin{equation}
\langle S_i^{-} S_i^{+} \rangle=\langle S_i^+ S_i^-\rangle 
= (2/3)\langle \vec{S}_{i}^2\rangle 
\;\;\;\; {\rm as}\;\;T\rightarrow T_{\rm N}^{-}.
\end{equation}

\section{Strong coupling limit}
We self consistently obtain the enhancement in transverse spin fluctuation 
$\langle S_i ^{-} S_i ^{+} \rangle $ 
due to thermal excitation of magnons.
The amplitude of the magnon propagator and the magnon energy are
both multiplicatively renormalized by $\langle S^z\rangle $,
the sublattice magnetization. 
This follows from a finite-temperature extension
of the one-loop correction to the magnon propagator, obtained earlier 
in the strong correlation limit.\cite{quantum} 
Alternatively, in the equivalent spin-$S$ Heisenberg model, this can be seen  within
the equation of motion approach,\cite{emm} which has been applied recently to the 
$S=1/2$ case of layered antiferromagnets.\cite{rps}
Considering first the simpler case of strong coupling, we therefore have
for the transverse spin fluctuation,
\begin{eqnarray}
& & \langle S_i ^{-} S_i ^{+} \rangle = \nonumber \\
& & \langle 2S^z \rangle 
\sum_q \left [\frac{1}{\sqrt{1-\gamma_q ^2}}\frac{1}{e^{\beta\omega_q} -1}
+\frac{1}{2}\left(\frac{1}{\sqrt{1-\gamma_q ^2}}-1\right )\right ] \; ,
\end{eqnarray}
where $\beta=1/T$ with $k_{\rm B}$ set to 1,
the magnon energy $\omega_q=zJ\langle S^z \rangle \sqrt{1-\gamma_q^2}$,
and the two terms are respectively the thermal and quantum (zero-point) contributions.
$\langle S_i^+ S_i^-\rangle$ is given by a similar equation except
with $[(1-\gamma_q^2)^{-1/2} + 1]$ for the quantum contribution. 
The renormalization of the magnon propagator by the magnitude 
$\langle 2S^z \rangle $ in Eq. (2) ensures that the  
commutation property of spin angular momentum 
operators $[S^+,S^-]=2S^z $ is obeyed by the 
expectation value in the spin state.
As  $\langle S^z\rangle$ vanishes in the limit
$T\rightarrow T_{\rm N}^{-}$,
from the isotropy requirement we obtain the N\'{e}el temperature,
\begin{equation}
T_{\rm N}=zJ\frac{S(S+1)}{3} \; f^{-1}_{\rm SF}\; ,
\end{equation}
where the spin-fluctuation correction factor 
$f_{\rm SF}\equiv \sum_q 1/(1-\gamma_q^2)$
in terms of $\gamma_q=\sum_{\mu=1}^{d}\cos q_\mu /d$.
For the simple cubic lattice $f_{\rm SF}=1.517$. 
As the mean-field result is $T_{\rm N}^{\rm MF}=zJS(S+1)/3$,
for the spin-$S$ Heisenberg model, spin fluctuations reduce $T_{\rm N}$ 
to nearly $70\%$ of its MF value. 
For $S=1/2$ this yields $T_{\rm N}=3.95t^2/U$, in good agreement with
the result $3.83t^2/U$ from high-temperature series 
expansions.\cite{high}
Furthermore, as the Goldstone mode is built-in, 
the spin fluctuation theory also yields the appropriate
$T^2$-falloff of the sublattice magnetization
at low temperatures $(T<<T_{\rm N})$, 
and the vanishing of $T_{\rm N}$ in
two dimensions, where $f_{\rm SF}$ diverges.

Interestingly, for hypercubic lattices 
this spin-fluctuation correction exactly matches with the 
Onsager-correction to the mean-field result.\cite{white}
Here one obtains the reaction-field correction factor 
$f_{\rm ORF}=\sum_q 1/(1+\gamma_q)$,
which is identically equal to the spin-fluctuation correction factor 
$f_{\rm SF}$ for hypercubic lattices.
It is thus quite a coincidence that 
while the (Onsager-corrected) MFT and SFT yield very different 
falloff of staggered magnetization with temperature,
identical N\'{e}el temperature is obtained in both theories for all $S$.

The Onsager correction has recently been utilized to discuss the finite-temperature
magnetism in the Hubbard model.\cite{logan} 
Here the low-energy excitations of the Hubbard model
are approximately mapped to those of an effective $S=1/2$ Heisenberg model 
with $U$-dependent, extended-range spin couplings.
The subsequent use of mean field theory, however, 
does not properly account for the continuous spin-rotational symmetry of the system,
resulting in an expression for $T_{\rm N}$ which is insensitive to whether 
the effective spin model is of the Heisenberg or Ising type.
As a vanishing N\'{e}el temperature would be obtained 
even for the Ising AF on a square lattice, 
the divergence of the reaction-field correction in two dimensions
should therefore be seen as a breakdown of the ORF theory in two dimensions,
rather than a source of the vanishing of $T_{\rm N}$. 
Consideration of the gapless spin-fluctuation modes 
is essential for this.

\section{General Case}
\subsection{Evaluation of transverse spin fluctuations ($T=0$)}
In order to extend these results to arbitrary $U/t$, we need the magnon
mode energies and spectral functions, and we now describe a new
method for evaluating transverse spin fluctuations at the RPA level,
from which both the N\'{e}el temperature and the quantum
spin-fluctuation correction to sublattice magnetization are obtained.
In addition to the pure antiferromagnets in two and three dimensions,
this approach is applicable to a variety of other systems such as 
AF with disorder, defects, vacancies etc., and provides a simple way 
to obtain first-order spin-fluctuation 
corrections about the HF-level broken-symmetry state.
The method is based on a convenient way to perform the frequency integral
to obtain spin correlations from spin propagators, and we illustrate it here 
for transverse spin correlations.
We write the time-ordered transverse spin propagator 
$\langle \Psi_{\rm G} | T [ S_i ^- (t) S_j ^+ (t')]|\Psi_{\rm G}\rangle$
at the RPA level in frequency space as,
\begin{equation}
\chi^{-+}(\omega)=\frac{\chi^0(\omega)}{1-U\chi^0(\omega)}
=\sum_n \frac{\lambda_n(\omega)}{1-U\lambda_n(\omega)} 
|\phi_n(\omega)\rangle \langle \phi_n(\omega) | \; ,
\end{equation}
where $\phi_n(\omega),\lambda_n(\omega)$ are the eigensolutions of the 
$\chi^0(\omega)$ matrix.
Here $\chi^0(\omega)$ is the zeroth-order, antiparallel-spin 
particle-hole propagator, evaluated in a suitable basis 
in the self-consistent, broken-symmetry state at the HF level.
In the site representation for instance,
$[\chi^0(\omega)]_{ij}=i\int (d\omega'/2\pi)
G_{ij}^{\uparrow}(\omega')G_{ji}^{\downarrow}(\omega'-\omega)$
in terms of the HF-level Green's functions.
Spin correlations are then obtained from,
\begin{eqnarray}
\langle S^- _i (t) S^+ _j (t')\rangle &=& -i\int \frac{d\omega}{2\pi}
[\chi^{-+}(\omega)]_{ij}\; e^{-i\omega (t-t')} \nonumber \\
&=& \pm \sum_n \frac{\phi_n ^i (\omega_n)\phi_n ^j (\omega_n)}
{U^2 (d\lambda_n/d\omega)_{\omega_n}} 
e^{-i\omega(t-t')} \; ,
\end{eqnarray}
where the collective mode energies $\omega_n$ are obtained from 
$1-U\lambda_n(\omega_n)=0$, and 
$\lambda_n(\omega)$ has been Taylor-expanded as 
$\lambda_n(\omega) \approx \lambda_n(\omega_n) + (\omega-\omega_n) 
(d\lambda_n/d\omega)_{\omega_n}$ near the mode energies
to obtain the residues.
For convergence, the retarded (advanced) part of the 
time-ordered propagator $\chi^{-+}$, having pole below (above) the 
real-$\omega$ axis, is to be taken for $t' < t$ ($t' > t$).
The frequency integral is conveniently replaced by an appropriate
contour integral in the lower or upper half-plane 
in the complex-$\omega$ space
for these two cases, respectively, which results in Eq. (5). 
 
For the pure AF case, it is convenient to use the two-sublattice representation
due to translational symmetry, and we work with 
the $2\times 2$ matrix $[\chi^0(q\omega)]$, 
which is given in terms of eigensolutions
of the HF Hamiltonian,\cite{spfluc} and numerically
evaluated using a momentum grid with $\Delta k=0.1$. Equal-time, same-site 
transverse spin correlations are then obtained from Eq. (5)
by summing over the different $q$ modes. 
We consider $t'\rightarrow t^-$, so that the retarded part is used, with
positive mode energies. 
From spin-sublattice symmetry, correlations on A and B sublattice sites 
are related via 
$\langle S^+ S^- \rangle_{A}= \langle S^- S^+ \rangle_{B}$.

\subsection{Quantum correction to sublattice magnetization}
From the commutation relation $[S^+,S^-]=2S^z$,  
the difference $\langle S^+ S^- - S^- S^+ \rangle_{\rm RPA}$
should yield 
$\langle 2S^z \rangle_{\rm HF}$, which is indeed confirmed
as shown in Fig. 1. The deviation at small $U$ is presumably because
of the finite grid size used in the calculations.
The sum $\langle S^+ S^- + S^- S^+ \rangle_{\rm RPA}$
yields a measure of transverse spin fluctuations about the HF state,
and for spin $S$ in the strong coupling limit, 
one obtains $\langle S^+ S^- + S^- S^+ \rangle_{\rm RPA}
=(2S)\sum_q 1/\sqrt{1-\gamma_q^2}$.\cite{quantum,magimp}
Using the identity,
$\langle S^z S^z\rangle =S(S+1) - \langle S^+ S^- + S^- S^+ \rangle/2$, 
the sublattice magnetization $m=\langle 2S^z\rangle$ 
is then obtained from,\cite{anderson}
\begin{equation}
\langle S^z\rangle =
S\left [1- \frac{1}{S} \left ( \frac{\langle S^+ S^- \rangle + \langle S^- S^+ \rangle}
{2S} -1 \right ) \right ]^{1/2} .
\end{equation}
To order $1/2S$, this yields the
correction to the sublattice magnetization 
of $(2S)^{-1}\sum_q [(1-\gamma_q^2)^{-1/2} -1]=0.156$ in three dimensions
for $S=1/2$. This same result at one-loop level
was earlier obtained from a different analysis in terms of
the electronic spectral-weight transfer,\cite{quantum} and 
is in exact agreement with the SWT result.\cite{anderson,oguchi}

For arbitrary $U$, as discussed above Eq. (1), 
$S\equiv \langle S^z\rangle_{\rm HF}$ plays the
role of the effective spin quantum number.
The sublattice magnetization $m$ is therefore obtained 
from $m=m_{\rm HF} -\delta m_{\rm SF}$, where 
the HF sublattice magnetization is obtained 
from the self-consistency condition,\cite{spfluc}
and the spin-fluctuation correction 
$\delta m_{\rm SF}$ at the one-loop level is obtained from
Eq. (6) with  $S=\langle S^z\rangle_{\rm HF}$,
\begin{equation}
\delta m_{\rm SF}=
\frac{\langle S^+ S^- +  S^- S^+ \rangle_{\rm RPA} }
{\langle S^+ S^- -  S^- S^+ \rangle_{\rm RPA} } -1 .
\end{equation}
The $U$-dependence of the sublattice magnetization $m(U)$ is shown in Fig. 1,
showing that it properly interpolates between the weak and strong coupling
limits, approaching the SWT result, $m(\infty)=m_{\rm SWT}=0.844$
as $U/t\rightarrow\infty$.
As expected, the spin-fluctuation correction $\delta m_{\rm SF}$ 
increases with interaction strength.
We note that the two low-$U$ datapoints 
are subject to the same finite-size numerical error
which affects the transverse spin correlations,
as mentioned earlier.

Earlier studies of $m(U)$ have employed
self-energy correction 
($d=2$, $m(\infty) \sim 0.7 \ne m_{\rm SWT}$),\cite{schrieffer}
the QMC method 
($d=2$),\cite{hirsch.tang}
functional-integral schemes
($d=3$, $m(\infty) = 1.0$),\cite{hasegawa2}
and recently an approximate mapping to an extended-range QHAF 
with effective $U$-dependent spin couplings obtained from the 
{\em static} transverse spin propagator at the RPA level
($d=3$, $m(\infty) = m_{\rm SWT}$).\cite{tusch_prb96}
In the last work, use of the static propagator implies that 
dynamical effects are not fully included,
which are in principle important in weak and intermediate 
coupling regimes when the magnon
energy is not negligible compared to the energy gap.

\subsection{Finite temperature}
To extend these calculations to finite temperature,
we write the diagonal elements 
$\chi^{-+}_{ii}(\omega)$ and  $\chi^{+-}_{ii}(\omega)$ in terms
of the RPA-level spectral functions and mode energies as,
\begin{equation}
\chi^{-+}_{ii}(\omega)=\sum_q \left [
\frac{-A_q}{\omega-\omega_q} 
+ \frac{B_q}{\omega+\omega_q}  \right ]   \; ,
\end{equation}
and a similar equation for 
$\chi^{+-}_{ii}(\omega)$ with $A_q$ and $B_q$ interchanged, 
the symmetry following from the identity
$\chi^{-+}(\omega)=\chi^{+-}(-\omega)$.
In the strong coupling limit, the spectral functions have the familiar form:
\begin{figure}
\vspace*{-70mm}
\hspace*{-28mm}
\psfig{file=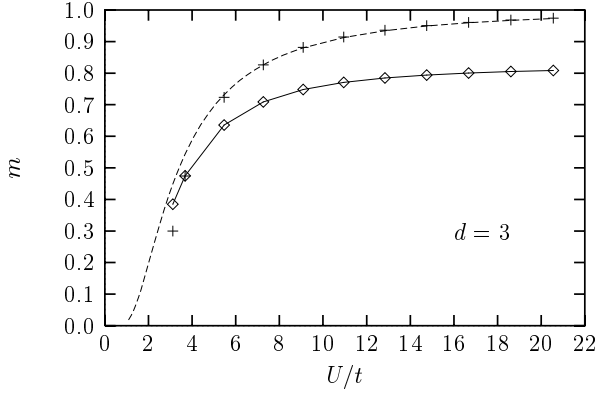,width=135mm,angle=0}
\vspace{-70mm}
\caption{The sublattice magnetization $m$ vs. $U$ (diamonds),
along with the HF results from (i)
the self-consistency condition (dashed), and (ii)
$\langle 2S^z\rangle_{\rm HF}=\langle S^+ S^- - S^- S^+\rangle_{\rm RPA}$
(plus).}
%\label{fig1}
\end{figure}

\noindent
\ \\
\ \\
$A_q=[(1-\gamma_q^2)^{-1/2} -1]/2$ and 
$B_q=[(1-\gamma_q^2)^{-1/2} +1]/2$. 
The transverse spin correlations at finite temperature are then obtained 
conventionally from these propagators 
by using the Poisson summation formulas which allow replacing the
sum over Matsubara frequencies by a contour integral in complex frequency space,
and by picking contributions from both poles in the propagators,
traversed in the clockwise direction.\cite{schrieffer.book} We obtain,
\begin{equation}
\langle S^- S^+\rangle =\sum_q \left [
\frac{-A_q}{e^{-\beta\omega_q}-1} 
+ \frac{B_q}{e^{\beta\omega_q}-1}  \right ]  \; .
\end{equation}
Both $A_q$ and $B_q$ are determined from the retarded part of $\chi^{-+}(\omega)$ 
having positive mode energies by using the spin-sublattice symmetry
$\chi_{AA}^{-+}(-\omega)=\chi_{AA}^{+-}(\omega)=\chi_{BB}^{-+}(\omega)$.
Thus the spectral functions are obtained from the magnon amplitudes on
A and B sublattices using,
\begin{equation}
A_q=\frac{(\phi_q^A)^2}
{U^2 (d\lambda_q/d\omega)_{\omega_q}} 
\; ; \;\;\;\; 
B_q=\frac{(\phi_q^B)^2}
{U^2 (d\lambda_q/d\omega)_{\omega_q}} \; .
\end{equation}

In analogy with the strong-coupling result 
discussed above Eq. (2), and in keeping with 
the renormalization constraints of this RPA theory
as discussed above Eq. (1),
we now multiplicatively renormalize both the magnon energy
$\omega_q$, and the magnon amplitudes $A_q,B_q$, 
by multiplying with the momentum-independent ratio
$\langle 2S^z\rangle/\langle 2S^z\rangle_{\rm HF}$.
This ensures that the commutation property $[S^+,S^-]=2S_z$
is satisfied by expectation values in the spin state,
as seen below.
By separating the zero- and finite-temperature 
contributions in $\langle S^- S^+\rangle$
and $\langle S^+ S^-\rangle$ from Eq. (9) and its counterpart,
it is seen that $\langle [S^+,S^-]\rangle
=\sum_q (B_q - A_q)=\langle 2S_z\rangle_{\rm HF}$ at all temperatures.
This HF result from use of the RPA-level transverse spin
correlations is corrected by the above renormalization, and 
the exact result $\langle 2S_z \rangle$ is obtained instead.
Furthermore, as in the strong coupling case, the renormalization 
of the magnon energy ensures that the spin stiffness
vanishes at the N\'{e}el temperature where long-range 
order vanishes. 
However, this does not constitute an independent renormalization as, 
in fact, the two renormalizations of magnon energy and amplitudes 
are connected. The only renormalization 
in $\chi^0(\omega)$ (of the $\omega$-term) which affects the magnon amplitudes
also modifies the magnon energy in precisely the same manner.\cite{quantum}

As $T\rightarrow T_{\rm N}^-$ and $\langle S^z\rangle\rightarrow 0$,
we therefore obtain from Eq.(9),
\begin{equation}
\langle S^- S^+\rangle =\langle S^+ S^-\rangle =
 T_{\rm N}\sum_q \left (\frac{\tilde{A}_q+\tilde{B}_q}
{\tilde{\omega}_q}\right ) \; ,
\end{equation}
where $\tilde{A}_q \equiv A_q/\langle 2S^z\rangle_{\rm HF}$,
$\tilde{B}_q \equiv B_q/\langle 2S^z\rangle_{\rm HF}$,
and ${\tilde{\omega}_q}\equiv \omega_q/\langle 2S_z \rangle_{\rm HF}$
are the scaled spectral functions and magnon energies.
As for the strong coupling case, $T_{\rm N}$ is obtained
from the isotropy requirement given in Eq. (1), with 
$\langle \vec{S}^2 \rangle = S(S+1)$ in terms of the
effective spin quantum number $S=\langle S_z\rangle_{\rm HF}
=m_{\rm HF}/2$. With this  
the N\'{e}el temperature is then obtained from
\begin{equation}
T_{\rm N}= \frac{2}{3} \langle \vec{S}^2 \rangle 
\left [ \sum_q \left ( \frac{\tilde{A}_q+\tilde{B}_q}
{\tilde{\omega}_q}\right ) \right ]^{-1} \; .
\end{equation} 
This result has the proper limiting behavior in both limits.
In the weak coupling limit, the magnon energy 
$\omega_q\approx 2\Delta$, the maximum magnon energy,
for almost all $q$ modes.
Also $\tilde{A}_q + \tilde{B}_q \approx 1$, so that
with $m_{\rm HF} \approx 0$, we obtain $T_{\rm N} 
\sim \Delta $, the proper HF scale. 
And in the strong coupling limit, 
$(\tilde{A}_q + \tilde{B}_q)/\tilde{\omega}_q = [3J(1-\gamma_q^2)]^{-1}$,
so that Eq. (3) is obtained, which is in good agreement with
the result from high-temperature series expansion, as discussed earlier.

Fig. 2 shows the variation of $T_{\rm N}$ with $U$, 
as obtained from above equation, 
along with the HF results 
which is asymptotically approached from below in weak coupling.
In strong coupling the result closely approaches the 
high-temperature series-expansion result $3.83t^2/U$,
which is well below the MFT result $6t^2/U$. 
For weak coupling HF-level finite-temperature
effects need to be considered in Eq. (12) when the temperature is
not negligible compared to the energy gap, and so 
it is the effective spin quantum number $S=m_{\rm HF}/2$ at
temperature $T_{\rm N}$ which is used here.
However, the magnon mode amplitudes and energies are
obtained at $T=0$.
Determination and use of magnon spectral properties at
temperature $T=T_{\rm N}$ is, of course, possible by
iteration, but not very practical. We find that
these HF-level finite-temperature corrections become
negligible for $U/t > 5$ in the crucial strong correlation limit. 
On the weak-coupling side of this, the energy gap 
and therefore the magnon energies $\omega_q$ would be 
reduced somewhat at finite temperature, so that  
the results for $T_{\rm N}$ are a slight overestimate
for $U/t < 5$.
\begin{figure}
\vspace*{-65mm}
\hspace*{-28mm}
\psfig{file=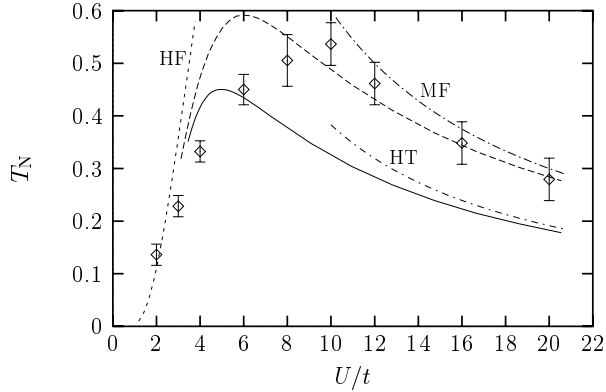,width=135mm,angle=0}
\vspace{-70mm}
\caption{N\'{e}el temperature $T_{\rm N}$ vs. $U/t$ (solid line), 
and using the dominant-mode approximation (dashed line).
Also shown are the strong-coupling asymptotes MF and HT,
from mean-field theory ($T_{\rm N}^{\rm MF}=6t^2/U$) 
and high-temperature series expansion 
($T_{\rm N}^{\rm HT}=3.83t^2/U$) respectively.}
\end{figure}

\subsection{Dominant-mode approximation}
We now consider a dominant-mode approximation which, 
in the strong coupling limit,
becomes equivalent to the dynamical mean-field theory. 
In the strong-coupling limit, 
where $\omega_q=dJ\sqrt{1-\gamma_q^2}$ in $d$ dimensions,
the magnon density of states 
is strongly peaked at the upper band edge at energy $dJ$, 
becoming a delta function in the limit $d\rightarrow\infty$.
As the highest-energy modes are dominant,
in this approximation we only consider these 
modes in the sum in Eq. (12).
As charge fluctuations which have been ignored 
are absent in the strong coupling limit,
and corrections to RPA  
decrease as $1/d$, this dominant-mode approximation
becomes exact in the limit of infinite dimensions. 
This approximation also becomes good in the weak-coupling limit,
where the magnon spectrum is strongly peaked at the gap energy $2\Delta$,
and all fluctuations are weak. 
Since for the highest-energy modes with $\omega_q=\omega_m$, 
we have $\tilde{A}_q +\tilde{B}_q \approx 1$, Eq. (12) simplifies to
\begin{equation}
T_{\rm N}=\frac{2}{3}\langle \vec{S}^2 \rangle \; \frac{\omega_m}{2S} \; ,
\end{equation}
which is also shown in Fig. 2. 
As $\omega_m=3J$ in the strong-coupling limit 
$U/t\rightarrow\infty$, this yields $T_{\rm N}=3J/2=6t^2/U$, the mean-field
result, which is also asymptotically  approached in the 
QMC\cite{hirsch,scalettar} and DMFT\cite{jarrell} calculations.
On the other hand, in the limit $U/t\rightarrow 0$, 
as $\omega_m$ approaches $2\Delta$, the energy gap, 
we obtain $T_{\rm N} \sim \Delta$, the proper weak-coupling scale. 
This analysis indicates that the essential features of the
$T_{\rm N}$ vs. $U$ behavior are contained 
in $\omega_m$, the energy scale of low-energy spin excitations
associated with spontaneous symmetry-breaking of the continuous
spin-rotational symmetry.
A relationship between $T_{\rm N}$ and the low-energy scale
was also obtained for the 
infinite-dimensional Bethe lattice.\cite{logan_jphys97}

At finite temperature, the gap parameter $\Delta$ is determined
from the self-consistency condition,
$1/U=\sum_k (1/2E_k)\tanh (\beta E_k/2)$, 
where $E_k=\sqrt{\Delta^2 +\epsilon_k^2}$ is the AF band energy. 
In the weak coupling limit, when the temperature is
comparable to the interaction strength, the gap
parameter, and therefore the maximum magnon energy, 
are sensitive to temperature.
We therefore use $\omega_m(U,T_{\rm N})$ in Eq. (13),
which is obtained from the magnon spectrum for a gap
parameter $\Delta(T_{\rm N})$ at temperature $T_{\rm N}$.
The self-consistent determination of $T_{\rm N}(U)$
is easily implemented in two steps ---
(i) obtaining $\omega_m(\Delta)$ and hence $T_{\rm N}(\Delta)$
as a function of the gap parameter $\Delta$, 
(ii) then using the self-consistency condition
to obtain $U(\Delta,T_{\rm N})$. 
As mentioned earlier, these HF-level finite-temperature
corrections become negligible for $U/t > 5$.
\begin{figure}
\vspace*{-65mm}
\hspace*{-28mm}
\psfig{file=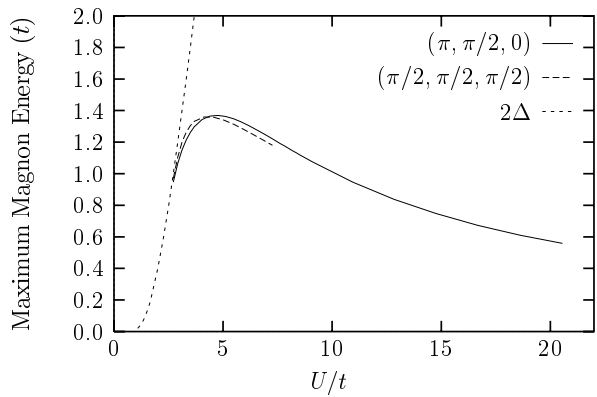,width=135mm,angle=0}
\vspace{-70mm}
\caption{Magnon energies $\omega_q$ vs. $U$ at momenta $(\pi,\pi/2,0)$ (solid)
and $(\pi/2,\pi/2,\pi/2)$ (dashed), 
together with the Hubbard gap $2\Delta$ (dotted).}
\end{figure}

The $U$-dependence of $\omega_m$ at $T=0$ is shown in Fig. 3, showing that it properly
interpolates between the two limits.
Similar qualitative nature of the U-dependence of the 
maximum magnon energy was found in two dimensions,\cite{spectrum}
and infinite dimensions.\cite{logan_prl96,logan_jphys97}
The complete spectrum
of spin excitations was studied earlier for the $d=2$ case.\cite{spectrum}
The band-edge degeneracy at energy $\omega_m=2J$ in the Heisenberg limit 
is removed for finite $U$, and the maximum was found to occur for
$q=(\pm \pi,0)$ and $(0,\pm \pi)$. For $d=3$ we find that
for most of the $U/t$ range the maximum occurs 
at momenta $q=(\pm\pi,\pm\pi/2,0)$ etc. where all the $\cos q_\mu$ terms
are completely different.
For small $U$ (less than about half bandwidth $B/2$) the magnon energy at
$(\pi,\pi/2,0)$ is actually not the maximum, as shown in Fig. 3.
However, the difference is marginal, and we have neglected this detail. 
As the degeneracy of $q$-points of the type $(\pi,\pi/2,0)$ is much higher, 
yielding a correspondingly higher DOS, 
it is reasonable to use this magnon energy as the 
relevant energy scale $\omega_m$. 
This also implies that the structure of the magnon density of states 
undergoes a crossover (at $U/t \approx 6$),
and for small $U$ the peak in magnon density of states 
changes position from the upper band edge to within the band.
\begin{figure}
\vspace*{-65mm}
\hspace*{-28mm}
\psfig{file=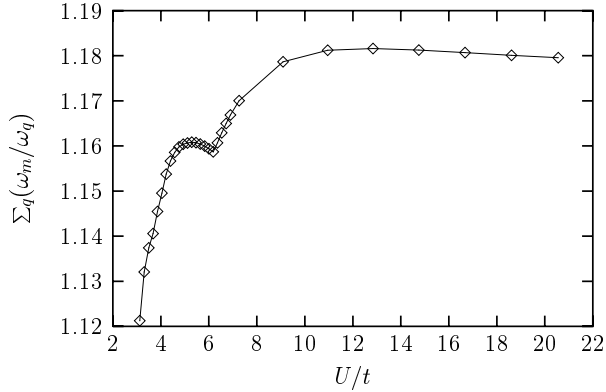,width=135mm,angle=0}
\vspace{-70mm}
\caption{The mode-averaged ratio $R\equiv \sum_q (\omega_m/\omega_q)$ vs. $U$,
showing the cusp at $U/t \approx 6$, which arises from 
the red-shift at lower-$U$ of the magnon DOS-peak energy relative
to the maximum magnon energy $\omega_m$.}
\end{figure}
\ \\
\ \\
This crossover has been seen in a recent study of the
$U$-dependence of the magnon density of states in $d=3$.\cite{tusch_prb96}

A convenient way to observe this crossover is 
through the $U$-dependence of the
mode-averaged ratio $R\equiv \sum_q (\omega_m/\omega_q)$.
Here $\omega_m$ is the maximum magnon energy (band edge), 
obtained by sampling through the full spectrum. 
$R$ approaches 1 in the limit when all magnon modes have 
energy approaching $\omega_m$. 
The $U$-dependence of $R$ is shown in Fig. 4.
With decreasing $U$ from the strong-coupling limit, 
there is at first a marginal increase in $R$, which then rapidly starts 
decreasing at $U \approx B$.
This decrease in $R$ 
indicates a reduction in the magnon dispersion, 
so that relatively more modes
have energy closer to the maximum magnon energy $\omega_m$.
This trend is, however,  sharply cutoff at $U \approx B/2=6t$,  
as clearly seen from the cusp in Fig. 4.
The sharp enhancement in $R$ with decreasing $U/t$ 
is due to the onset of a red-shift of the DOS-peak energy
(where the magnon DOS peaks) relative to the band edge, $\omega_m$.
Because of this red-shift
most magnon modes  have energy less than $\omega_m$, 
yielding an enhanced contribution to $R$.
The rapid decrease in $R$ towards 
unity with further decrease in $U$ 
indicates that almost all magnon modes have energy 
nearly equal to $\omega_m$ which, as mentioned earlier, 
approaches the charge gap $2\Delta$  in weak coupling.
The role of the approaching Stoner band in
compressing the magnon spectrum has been noted earlier.\cite{spectrum}

In conclusion, this analysis in terms of transverse spin fluctuations
at the RPA level provides a good account of the magnetic phase boundary 
for the three-dimensional, half-filled Hubbard model,
particularly in the crucial strong coupling limit.
The quantum correction to sublattice magnetization as well as 
the N\'{e}el temperature are obtained in the whole $U/t$ range,
and both interpolate properly 
between the weak and strong coupling limits.
This analysis and the new method to evaluate 
transverse spin fluctuations 
should also prove valuable to other related systems such as
antiferromagnets with disorder, defects and vacancies, where
spin-fluctuation studies can be extended to strong disorder and
strong defect/vacancy concentrations.

\section*{ACKNOWLEDGMENTS} 
Helpful conversations with D. Vollhardt and 
M. Ulmke, and support from the Alexander von Humboldt 
Foundation through a Research Fellowship
are gratefully acknowledged.


\newpage
\begin{references}

\bibitem[*]{asingh}On leave from Department of Physics, Indian Institute of
Technology, Kanpur 208016.

\bibitem[1]{hasegawa}H. Hasegawa, J. Phys. Soc. Jpn. {\bf 49}, 178 (1980).

\bibitem[2]{kakehashi}Y. Kakehashi and J. H. Samson, Phys. Rev. B {\bf 33}, 298 (1986).

\bibitem[3]{kakehashi2}Y. Kakehashi and H. Hasegawa, Phys. Rev. B {\bf 36},
4066 (1987); {\bf 37}, 7777 (1988).

\bibitem[4]{hasegawa2}H. Hasegawa, J. Phys. Condens. Matter. {\bf 1}, 9325 (1989).

\bibitem[5]{moriya}{\em Electron Correlation and Magnetism in Narrow-Band
Systems}, edited by T. Moriya (Springer-Verlag, Berlin, 1981).

\bibitem[6]{moriya2}T. Moriya, {\em Spin Fluctuations in Itinerant Electron 
Magnetism} (Springer-Verlag, Berlin, 1985).

\bibitem[7]{hirsch}J. E. Hirsch, Phys. Rev. B {\bf 35}, 1851 (1987).

\bibitem[8]{scalettar}R. T. Scalettar, D. J. Scalapino,
R. L. Sugar, and D. Troussaint, Phys. Rev. B {\bf 39}, 4711 (1989).

\bibitem[9]{jarrell}M Jarrell, Phys. Rev. Lett. {\bf 69}, 168 (1992);
For a review see A. Georges, G. Kotliar, W. Krauth, and M. Rozenberg,
Rev. Mod. Phys. {\bf 68}, 13 (1996).

\bibitem[10]{vollhardt}W. Metzner and D. Vollhardt, Phys. Rev. Lett. {\bf 62},
324 (1989).

\bibitem[11]{comparison}For a recent compilation of earlier results, see reference [12].

\bibitem[12]{logan}Y. H. Szczech, M. A. Tusch, and D. E. Logan,
Phys. Rev. Lett. {\bf 74}, 2804 (1995).

\bibitem[13]{quantum}A. Singh, Phys. Rev. B {\bf 43}, 3617 (1991).

\bibitem[14]{emm}S. Doniach and E. H. Sondheimer, {\em Green's Functions for 
Solid State Physicists}, Benjamin/Cummings (1974).

\bibitem[15]{rps}R. P. Singh and M. Singh, Phys. Stat. Sol. (b) {\bf 169}, 571 (1992).

\bibitem[16]{high}G. S. Rushbrooke, G. A. Baker jr., and P. J. Wood,
in {\em Phase Transitions and Critical Phenomena}, 
Vol. 3, Chapter 5 (Academic, New York, 1974).

\bibitem[17]{white}See, e.g., R. M. White, {\em Quantum Theory of Magnetism}
(Springer-Verlag, Berlin, 1983).

\bibitem[18]{spfluc}A.Singh and Z. Te\v{s}anovi\'{c},
Phys. Rev. B {\bf 41}, 614 (1990); Phys. Rev. B {\bf 41}, 11 457 (1990).

\bibitem[19]{magimp}A. Singh and P. Sen, Phys. Rev. B {\bf 57}, 10598 (1998). 

\bibitem[20]{anderson}P. W. Anderson, Phys. Rev. {\bf 86}, 694 (1952).

\bibitem[21]{oguchi}T. Oguchi, Phys. Rev. {\bf 117}, 117 (1960).

\bibitem[22]{schrieffer}J. R. Schrieffer, X.-G. Wen, and S.-C. Zhang,
Phys. Rev. B {\bf 39}, 11663 (1989).

\bibitem[23]{hirsch.tang}J. E. Hirsch and S. Tang, 
Phys. Rev. Lett. {\bf 62}, 591 (1989).

\bibitem[24]{tusch_prb96}M. A. Tusch, Y. H. Szczech, and D. E. Logan,
Phys. Rev. B {\bf 53}, 5505 (1996).   

\bibitem[25]{schrieffer.book}See, e.g., R. J. Schrieffer, {\em Theory of
Superconductivity} (Addison-Wesley, Reading, 1964).

\bibitem[26]{logan_jphys97}Y. H. Szczech, M. A. Tusch, and D. E. Logan,
J. Phys. Cond. Matt. {\bf 9}, 9621 (1997).

\bibitem[27]{spectrum}A. Singh, Phys. Rev. B {\bf 48}, 6668 (1993).

\bibitem[28]{logan_prl96}D. E. Logan, M. P. Eastwood, and M. A. Tusch,
Phys. Rev. Lett. {\bf 76}, 4785 (1996).

\end{references}
\end{document}